\documentclass[preprint,aps,showpacs,preprintnumbers,amsmath,amssymb,nofootinbib]
{revtex4}
\usepackage{epsfig}
\usepackage{axodraw}
\begin{document}

\begin{flushright}
\end{flushright}

\newcommand{\be}{\begin{equation}}
\newcommand{\ee}{\end{equation}}
\newcommand{\bea}{\begin{eqnarray}}
\newcommand{\eea}{\end{eqnarray}}
\newcommand{\nn}{\nonumber}

\def\lb{\Lambda_b}
\def\ll{\Lambda}
\def\mb{m_{\Lambda_b}}
\def\ml{m_\Lambda}
\def\s1{\hat s}
\def\ds{\displaystyle}


\title{\large Effect of FCNC mediated $Z$ boson on lepton flavor violating decays}
\author{Rukmani Mohanta }
\affiliation{ School of Physics, University of Hyderabad, Hyderabad
- 500 046, India}

\begin{abstract}
We study the three body lepton flavor violating (LFV) decays  $\mu^-
\to e^- e^+ e^-$,  $\tau^- \to l_i^- l_j^+ l_j^-$ and the
semileptonic decay $\tau \to \mu \phi $ in the  flavor changing
neutral current (FCNC) mediated $Z$ boson model. We also calculate
the branching ratios for LFV leptonic $B$ decays, $B_{d,s} \to \mu
e$, $B_{d,s} \to \tau e$, $B_{d,s} \to \tau \mu$ and the conversion of muon 
to electron in Ti nucleus.  The new
physics parameter space is constrained by using the experimental
limits on $\mu^- \to e^- e^+ e^-$ and  $\tau^- \to \mu^- \mu^+
\mu^-$. We find that the branching ratios for $\tau \to eee $ and
$\tau \to \mu \phi$ processes could be as large as $\sim {\cal
O}(10^{-8})$ and ${\rm Br} (B_{d,s} \to \tau \mu,~ \tau e) \sim
{\cal O}(10^{-10})$. For other LFV $B$ decays the branching ratios
are found to be too small to be observed in the near future.

\end{abstract}

 \maketitle

\section{Introduction}

It is very well known that in the standard model (SM) of electroweak
interactions the generation lepton number is exactly conserved.
However, the observation of neutrino oscillation implies that family
lepton number must be violated.  The neutrino oscillation is due to
a mismatch between the weak and mass eigenstates of neutrinos and
this mismatch causes mixing between different generation of leptons
in the charged current interaction of the $W$ boson. Due to the the
violation of family lepton number, flavour changing neutral current
(FCNC) processes in the lepton sector could in principle occur,
analogous to the  quark sector. Some examples of FCNC transitions in
the lepton sector would be $l_i \to l_j \gamma$, $l_i \to l_j l_k
\bar l_k$, $B \to l_i \bar l_j$ and $B \to X_s l_i \bar l_j$ (where
$l$ is any charged lepton) etc. Although there is no direct
conclusive  experimental evidence for such processes that have been
observed so far, but there exist severe constraints on some of these
LFV decay modes \cite{pdg}. It should be noted that FCNC transitions
in the lepton sector that are solely due to mixing in the charged
current interaction with the usual left handed $W$ boson and light
neutrinos are extremely small because they are suppressed by powers
of $m_\nu^2/M_W^2$. In particular the branching ratio for $\mu \to e
\gamma$ in the SM amounts to at most $10^{-54}$ \cite{buras1}, to be
compared with the present experimental upper bound $1.2 \times
10^{-11}$ \cite{pdg}. Therefore any observation of lepton flavour
violation (LFV) in the foreseeable future would be an unambiguous
signal of new physics beyond the SM. One way to increase the FCNC
interactions in the lepton sector is to introduce heavy neutrinos so
that the suppression factor $m_\nu^2/M_W^2$ is not affected. This
can be done for example by introducing heavy fourth generation
\cite{buras2}. If one insists on having just three left handed
neutrinos one needs to give the right handed neutrinos heavy
Majorana masses. As a consequence, these observations often put
severe constraints on the parameter space of new physics models in
which heavy leptons are present. Moreover these decays being
unaffected by hadronic uncertainties, allow for a clear distinction
between different new physics scenarios, in particular when several
branching ratios are considered simultaneously.

Many models for physics beyond the standard model predicts lepton
flavor violating decays of charged leptons at a level which may
become observable very soon \cite{tau}. The LFV tau decays are
analyzed in a model independent way in Ref. \cite{mannel}.
 In the present paper we investigate the
LFV $\mu$ and  $\tau$ decays e.g., $\mu, \tau \to l_i l_j \bar l_j
$, $\mu \to e \nu_e \bar \nu_\mu$, $\tau \to \mu \phi$ and $B_{d,s}
\to l_i \bar l_j$ in a model where $Z$ mediated flavor changing
neutral current (FCNC) transitions occur at the tree level. It is
well known that FCNC coupling of the $Z$ boson can be generated at
the tree level in various exotic scenarios. Two popular examples
discussed in the literature are the models with an extra $U(1)$
symmetry \cite{ref2} and those with the addition of non-sequential
generation of quarks or leptons \cite{ref3}. In the case of extra
$U(1)$ symmetry the FCNC couplings of the $Z$ boson are induced by
$Z - Z^\prime$ mixing, provided the SM quarks/leptons have family
non-universal charges under the new $U(1)$ group. In the second
case, adding a different number of quarks or leptons, the pseudo
mixing matrix needed to diagonalize the charged currents is no
longer unitary and this leads to tree level FCNC couplings. It
should be noted that, recently, there has been renewed interests
shown in the literature concerning the non-universal $Z$ induced new
physics \cite{barger} in the quark sector.

Here we will follow the second approach \cite{ref3} to analyze some
FCNC induced rare LFV decays. We consider the presence of an
additional vector like sterile neutrino, which could mix with the SM
three neutrinos resulting a  $4 \times 4$ mixing matrix  $V$ for the
neutrinos. Due to such mixing however  the charged current
interactions remain unchanged except that the SM PMNS mixing matrix
$V_{PMNS}$ is now the $3 \times 4$ upper sub-matrix of $V$. However,
the distinctive feature of this model is that the FCNC interaction
enters the neutral current Lagrangian of the left handed neutrinos as
\be {\cal L}_Z= \frac{g}{2 \cos \theta_W}\Big[\bar l_{Li} \gamma^\mu
l_{Li}-\bar \nu_{L \alpha}U_{\alpha \beta} \gamma^\mu \nu_{L
\beta}-2 \sin^2 \theta_W J^\mu_{em}\Big] Z_\mu\;, \ee with \be
U_{\alpha \beta}= \sum_{i= \nu_e, \nu_\mu, \nu_\tau, \nu_s}
V_{\alpha i}^\dagger V_{i \beta}= \delta_{\alpha \beta}-V_{4
\alpha}^* V_{4 \beta}\;, \ee where $U$ is the neutral current mixing
matrix for the neutrino sector, which is given above. As $V$ is not
unitary, $U \neq {\bf 1}$. In particular the non-diagonal elements
do not vanish. \be U_{\alpha \beta}=-V_{4 \alpha}^* V_{4 \beta} \neq
0 ~~~~ {\rm for}~~~\alpha \neq \beta\;. \ee Since the various
$U_{\alpha \beta}$ are non vanishing, they would signal new physics
and the presence of FCNC at the tree level and this can
substantially mediate many low energy LFV processes. In this paper
we consider the impact of FCNC mediated $Z$ boson couplings on
several  LFV decays of $\mu$, $\tau$ and $B_{d,s}$ mesons. There is
also huge experimental efforts going on to look for any possible
signals of LFV decays. For instance, the recent commencement of the
MEG experiment \cite{meg}, which will probe ${\rm Br}(\mu \to e
\gamma) \sim 10^{-13}$ two orders magnitude beyond the current
limit. Also the $B$ factories, Belle and Babar have searched for the
decay modes $ \tau \to l_i l_j \bar l_j$ with upper limits in the
range ${\rm Br}( \tau \to l_i l_j \bar l_j) < (2-8) \times 10^{-8}$
\cite{bfac}. Searches for $\tau \to \mu \mu \bar \mu$ can be
performed at the Large Hadron Collider where $\tau$ leptons are
copiously produced from the decays of $W$, $Z$, $B$ and $D$, with
anticipated sensitivities ${\rm Br}(\tau \to \mu \mu \bar \mu) \sim
10^{-8}$ \cite{lhc}.

The outline of the paper is as follows. The LFV decays $\mu^- \to e^-
e^+ e^-$  and $\mu^- \to e^- \nu_e \bar \nu_\mu$ are presented in section II
and $B_{d,s} \to \mu^\pm e^\mp$ in section III. The LFV tau decays are
discussed in section IV. In section V we discuss the
$\mu -e$ conversion in Titanium and $\mu \to e \gamma $ process.
section VI contains the summary and conclusion.

\section{Decay rates for $\mu^- \to e^- e^+ e^- $ and $\mu^- \to e^- \nu_e \bar \nu_\mu$}

Let us now consider the decay mode $\mu^-(p) \to e^-(p_1) + e^-(p_2)
+e^+(p_3) $ which has a strict bound of $10^{-12}$. Since this mode
is highly suppressed in the SM with a branching ratio ${\rm Br}(\mu
\to eee)\simeq 10^{-17}$ \cite{musm}, where the presence of a
right-handed neutrino is assumed. Therefore, in our analysis we will
not include the SM contributions to this process. In the FCNC
mediated  $Z$ boson model, it occurs at the tree level due to the
presence of FCNC coupling of the $Z$ boson.
 Hence it is expected that in such a model the branching ratio can
be substantially enhanced and it could be possible that this mode
can be probed at current and forthcoming experiments. The  Feynman
diagram for this process is shown in Figure 1, where the blob
represents the tree level FCNC coupling of $Z$ boson (lepton flavor
violating coupling). Thus, one can obtain the amplitude for this
process as \be \frac{G_F}{\sqrt 2}U_{\mu e}\Big ( [\bar e(p_1)
\gamma^\mu (1- \gamma_5) \mu(p)][\bar e(p_2) \gamma_{\mu}(C_V-
C_A\gamma_5) e(p_3)] +(p_1 \leftrightarrow p_2)\Big ), \ee where
$p_i$'s represent the momenta of different particles involved in
this process. $C_V$ and $C_A$ are the vector and axial-vector
couplings of the $Z$ boson to electron-positron pair, with values
$C_V = -1/2 + 2 \sin^2 \theta_W$ and $C_A=-1/2$. $U_{\mu e}$ is the
lepton flavor violating FCNC coupling at the $\mu e Z$ vertex.

Thus, from the above amplitude, one can obtain the  decay rate after
doing a simple calculation and three body phase space integration, as
\be \Gamma (\mu^- \to e^- + e^+ + e^- ) = \frac{G_F^2}{384 \pi^3}
|U_{\mu e}|^2 (C_A -C_V)^2 m_\mu^5\;, \ee where we have neglected
the electron mass. Now using the experimental upper limit of the branching
ratio of this mode ${\rm Br}(\mu^- \to e^- e^+ e^-) < 10^{-12}$
\cite{pdg}, we obtain the upper limit on the lepton flavor violating
coupling $U_{\mu e}$ as \be |U_{\mu e}| < 3.05 \times 10^{-6}, \ee
where we have used the mass and lifetime of muon from \cite{pdg} and
$\sin^2 \theta_W=0.231$.

\hspace*{1.0 truein}
\begin{figure}[t]
\begin{center}
\begin{picture}(300,50)(0,0)
\SetColor{Black}
\Photon(100,25)(175,25){5}{6}
\SetColor{Black}
\ArrowLine(50,0)(100,25){}{}
\SetColor{Black}
\ArrowLine(100,25)(50,55){}{}\Vertex(100,25){2}
\SetColor{Black}
\ArrowLine(225,0)(175,25){}{}
\SetColor{Black}
\ArrowLine(175,25)(225,55){}{}

\vspace*{0.8 true in}
\Text(135,40)[]{$Z$}
\Text(75,50)[]{$e^-$}
\Text(75,05)[]{$\mu^-$}
\Text(205,55)[]{$e^-$}
\Text(205,-02)[]{$e^+$}

\end{picture}
\end{center}
\caption{Feynman diagram for $\mu^- \to e^- e^+ e^-$ in the model
with FCNC mediated $Z$ boson, where the blob represents the tree
level FCNC coupling (lepton flavor changing vertex) of $Z$ boson.}
\end{figure}
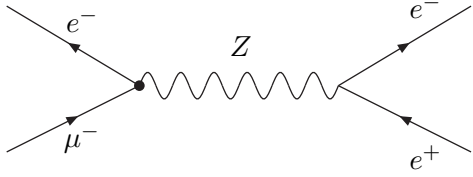

Next we consider another LFV $\mu$ decay, $\mu^- \to e^- \nu_e \bar
\nu_\mu$, which violates $L_e$ and $L_\mu$ by two units each and
hence it is highly suppressed in the SM. However, in the model with
FCNC mediated $Z$ boson this process can occur at the tree level
with two LFV vertices  and the corresponding Feynman diagram is
shown in Figure-2 where $l_1$ and $l_2$ denote $\mu$ and $e$,  $f_1$ and
$f_2$ as $\nu_e$ and $\nu_\mu$. The
amplitude for this process is given as \bea {\cal M}(\mu^- \to e^-
\nu_e \bar \nu_\mu)= \frac{G_F}{2 \sqrt 2} |U_{\mu e}|^2 [\bar e
\gamma^\mu (1-\gamma_5) \mu ][\bar \nu_{\mu} \gamma_\mu (1-\gamma_5)
\nu_e]\;. \eea

The corresponding decay rate is found to be (neglecting
electron mass) \bea \Gamma = \frac{G_F^2}{384 \pi^3} |U_{\mu e}|^4
m_{\mu}^5 .\label{brmu}\eea  \hspace*{1.0 truein}
\begin{figure}[b]
\begin{center}
\begin{picture}(300,50)(0,0)
\SetColor{Black}
\Photon(100,25)(175,25){5}{6}
\SetColor{Black}
\ArrowLine(50,0)(100,25){}{}
\SetColor{Black}
\ArrowLine(100,25)(50,55){}{}\Vertex(100,25){2}
\SetColor{Black}
\ArrowLine(225,0)(175,25){}{}\Vertex(175,25){2}
\SetColor{Black}
\ArrowLine(175,25)(225,55){}{}

\vspace*{0.8 true in}
\Text(135,40)[]{$Z$}
\Text(75,50)[]{$l_2$}
\Text(75,05)[]{$l_1$}
\Text(205,55)[]{$f_1$}
\Text(205,-02)[]{$\bar f_2$}

\end{picture}
\end{center}
\caption{Feynman diagram
 for $l_1 \to l_2 f_1 \bar f_2$ in the model with FCNC mediated
 $Z$ boson, where the blobs represent the tree level FCNC
vertices.}
\end{figure}
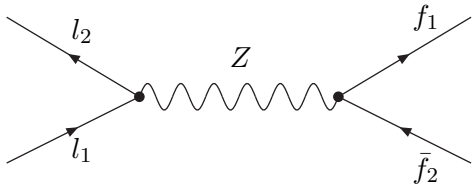

Now varying the value of $|U_{\mu e}|$ between $0\leq |U_{\mu e}
\leq 3.05 \times 10^{-6}$, extracted from  $\mu \to e e e$ process,
we show in Figure-3 the correlation plot between the branching
ratios of $\mu^- \to e^- e^+ e^-$ and $\mu^- \to e^- \nu_e \bar
\nu_\mu $ processes. From the figure it can be seen that the maximum
value of branching ratio that can be accommodated for the process
$\mu^- \to e^- \nu_e \bar \nu_\mu$ in the extra $Z$ boson model
considered here as \bea {\rm Br}(\mu^- \to e^- \nu_e \bar \nu_\mu )
< 4.36 \times 10^{-23}\;, \eea which is well below the present
experimental upper limit \cite{pdg} \bea {\rm Br}(\mu^- \to e^- \nu_e \bar
\nu_\mu ) < 1.2 \times 10^{-2}. \eea Since the branching ratio is
well below the sensitivities of the present and upcoming experiments
there is no chance to observe this mode in the near future.
Since the final neutrinos are difficult to detect it is useful to
consider the channel $\mu^- \to e^- \nu_\alpha \bar \nu_\alpha $,
which involves only one FCNC $Z-e-\mu$ coupling in contrast to
the previous case where there are two such couplings. The decay rate
for this process is given as
\bea \Gamma(\mu \to e \nu_\alpha \bar \nu_\alpha) = \frac{G_F^2}{384 \pi^3} |U_{\mu e}|^2
m_{\mu}^5. \eea
Using the bound on $|U_{\mu e}|$ and summing over all neutrino flavors,
 we obtain the branching ratio as
\be
{\rm Br}(\mu \to e \nu \bar \nu) < 1.4 \times 10^{-11},
\ee
which could be accessible in the future high sensitivity experiments.

 \begin{figure}[htb]
  \centerline{\epsfysize 2.25 truein \epsfbox{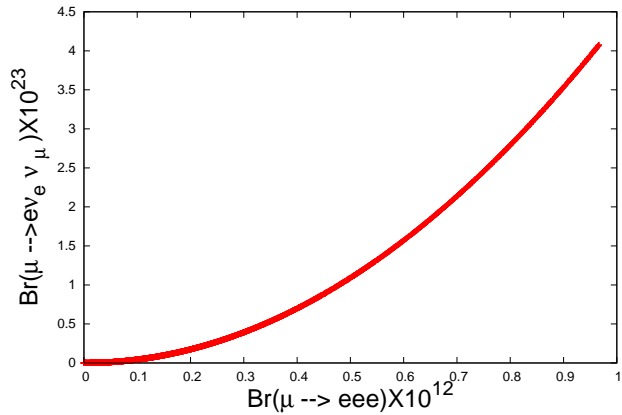}}
\caption{Correlation plot between the branching ratios of $\mu^- \to
e^- e^+ e^- $ and $\mu^- \to e^- \nu_e \bar \nu_\mu $ processes.}
  \end{figure}
\section{$B_{d,s} \to \mu^\pm e^\mp $}
Now we consider the lepton flavour violating $B$ meson decays
$B_{d,s} \to l_i \bar l_j$. Here we will consider only the decay
mode $B_s \to \mu^\pm e^\mp$, since the same formula will hold good
for all the above mentioned processes by replacing appropriate
particle masses.  The corresponding Feynman diagram will be similar
to the figure-2, with two FCNC vertices one for the quark and one
for the lepton parts i.e., replacing $l_1$ and $l_2$ by $b$ and $s$
and ($f_i$, $f_2$) by ($\mu$, $e$). The effective Hamiltonian
describing this process is given by \be {\cal H}_{eff} =
\frac{G_F}{2 \sqrt 2} U_{bs} U_{\mu e} (\bar s
\gamma^\mu(1-\gamma_5) b )(\bar \mu \gamma_\mu (1-\gamma_5) e)\;,
\ee where  $U_{bs}$ represents the FCNC coupling of the quark
sector.  To evaluate the transition amplitude, we use the following
matrix element of the quark current between the initial $B_s$ meson
and $vacuum$ as \be \langle 0 | \bar s \gamma^\mu \gamma_5 b |
B_s(p_B) \rangle = i f_{B_s} p_B^\mu\;. \ee Thus, we obtain the
amplitude for $B_s \to \mu e$ process as \bea {\cal M}(B_s \to \mu
e) = \frac{-i G_F}{2 \sqrt 2} U_{bs} U_{\mu e} m_\mu f_{B_s} [\bar
\mu (1-\gamma_5) e], \eea and the corresponding decay width as \bea
\Gamma(B_s \to \mu^\pm e^\mp) = \frac{G_F^2}{16 \pi} m_{\mu}^2
m_{B_s} |U_{sb} U_{\mu e}|^2 f_{B_s}^2 \left (1
-\frac{m_{\mu}^2}{m_B^2} \right)^2\;. \eea For numerical estimation
we use the particle masses and lifetime from \cite{pdg}, the $B_s$
meson decay constant as $f_{B_s}=0.24$ GeV. Now using the value of
$|U_{bs}|$ as $0 \leq |U_{bs}| \leq 0.005$ \cite{rm1}, which is
extracted from the mass difference of $B_s - \bar B_s$ system and
varying $|U_{\mu e}|$ between $0\leq |U_{\mu e}|< 3.05 \times
10^{-6}$, we show in figure-4 the correlation plot between the
branching ratios of $\mu^-\to e^- e^+ e^-$ and $B_s \to \mu^\pm
e^\mp$. From the figure, one can obtain the upper limit of the
branching ratio as \bea {\rm Br}(B_s \to \mu^\pm e^\mp) < 4.7 \times
10^{-18}. \eea Similarly for $B_d \to \mu^\pm e^\mp$ process using
the value of FCNC $Zbd$ coupling as $|U_{bd}|\leq 10^{-3}$
\cite{ubd} and $f_{B_d}=0.22$ GeV, we obtain the corresponding
branching ratio upper limit as \bea {\rm Br}(B_d \to \mu^\pm e^\mp)
< 1.7 \times 10^{-19}. \eea Since these rates are also highly
suppressed there is also no possibility to observe these modes in
the near future.
\begin{figure}[htb]
\centerline{\epsfysize 2.25 truein \epsfbox{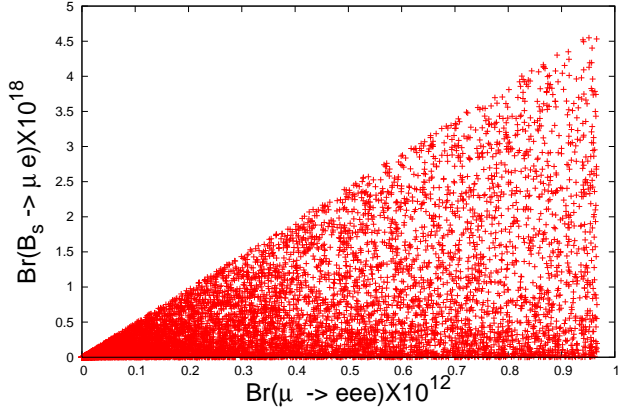}}
\caption{Correlation plot between the branching ratios of $\mu^- \to
e^- e^+ e^- $ and $B_s \to \mu^\pm e^\mp$ processes.}
\end{figure}
\section{lepton flavour violating $\tau $ decays }
Now we consider the lepton flavour violating $\tau$ decays. The LFV
decays $\tau \to lll$ are in analogy of $\mu \to eee$ and provide
sensitive probe of the lepton flavor violating couplings $U_{\tau
l}$ in the FCNC mediated $Z$ boson model. More observation of such
decays would constitute a spectacular signal of physics beyond the
SM. There are six distinct decays for $\tau^- \to lll $: $\tau^- \to
\mu^- \mu^+ \mu^-$, $\tau^- \to e^- e^+ e^-$, $ \tau^- \to \mu^-
\mu^+ e^-$, $\tau^- \to \mu^- \mu^- e^+ $, $\tau^- \to e^- e^+
\mu^-$, $\tau^- \to e^- e^- \mu^+$. Searches for all six decays have
been performed by BABAR  and Belle \cite{bfac} and upper limits of
the order ${\rm Br}(\tau \to lll)\sim {\cal O}( 10^{-8})$ are
obtained. Although these limits are several orders of magnitude
weaker than the bound ${\rm Br}(\mu \to eee) < 10^{-12}$, they have
the virtue of constraining many combinations of $U_{ij}$. Moreover
greater sensitivity to ${\rm Br}(\tau \to lll)$ is expected from
forthcoming experiments.

Here we will only focus on $\tau \to l_i l_j \bar l_j$ processes
having only one LFV vertex. The corresponding Feynman diagram will
be analogous to that of Figure-1 describing $\mu \to eee$ process.
First we will focus on $\tau^- \to \mu^- \mu^+ \mu^-$  and $\tau^-
\to e^- e^+ e^-$ which will
allow us to obtain the constraint on $U_{\tau \mu} $ and $U_{\tau e}$.
The branching ratios for such processes  will have the same form as
Eq. (\ref{brmu}) with the muon
mass replaced by the $tau$ mass and taking  into account the
appropriate LFV coupling $U_{\tau l}$. Using the upper limits of the
branching ratios from \cite{pdg}, we obtain the limits on the LFV couplings as
follows. \bea {\rm Br}(\tau^- \to e^- e^+ e^-) < 3.6 \times 10^{-8},
\Rightarrow |U_{\tau e}|<1.37 \times 10^{-3}\nn\\
{\rm Br}(\tau^- \to \mu^- \mu^+ \mu^-) < 3.2 \times 10^{-8}
\Rightarrow |U_{\tau \mu}|<1.295 \times 10^{-3}. \eea

These constraints can be used to predict the branching ratios of
LFV $B$ decays $B_{d,s} \to \tau \mu , \tau e$.
Now using the above constraint on $U_{\tau \mu}$,  in Figure-5 we
show the correlation plot between the branching ratios of $\tau^- \to
\mu^- \mu^+ \mu^-$ and $B_s \to \tau \mu$. The branching ratio upper limit
for this process is found to be
\bea
{\rm Br}(B_s \to \tau \mu)=1.9 \times 10^{-10}.
\eea

 \begin{figure}[htb]
  \centerline{\epsfysize 2.25 truein \epsfbox{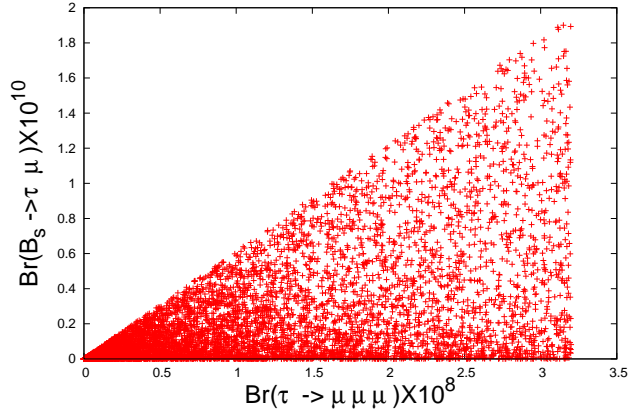}}
\caption{Correlation plot between the branching ratios of $\tau^-
\to \mu^- \mu^+ \mu^- $ and $B_s \to \tau \mu $ processes.}
  \end{figure}

\subsection{$\tau^- \to e^- \mu^+ \mu^- $}
Next we consider the decay process $\tau^- \to e^- \mu^+ \mu^-$. The Feynman
diagram for this process is similar to Fig-1  with one LFV coupling. Although
this process can also have contribution with two LFV couplings but such contribution is
highly suppressed. The amplitude for this process is given as \be {\cal M}(\tau^- \to
e^- \mu^+ \mu^-) = \frac{G_F}{\sqrt 2} U_{\tau e}[\bar e(p_1)
\gamma^\mu (1- \gamma_5) \tau(p)][\bar \mu(p_2) \gamma_\mu(C_V - C_A
\gamma_5) \mu(p_3)],
\ee
 and the corresponding branching ratio as
 \be \Gamma(\tau^- \to e^- \mu^+ \mu^-)=\frac{G_F^2}{384
\pi^3} |U_{\tau e}|^2 m_\tau^5(C_V^2+C_A^2). \ee

Using the branching ratio $Br(\tau^- \to e^- e^+ e^-) \leq 3.7 \times
10^{-8}$ we obtain the constraint on $|U_{\tau e}|$ as \be|U_{\tau
e}|< 1.28 \times 10^{-3}. \ee

\subsection{$\tau \to \mu \phi $ }

Here we consider the LFV violating  semileptonic decay mode of
$\tau$ lepton $\tau \to \mu \phi$ and the corresponding Feynman
diagram is analogous to Fig-1. The amplitude for this process \bea
{\cal H}_{eff} = \frac{G_F}{\sqrt 2} U_{\tau \mu} (\bar \mu
\gamma^\mu (1-\gamma_5) \tau)
 (\bar s \gamma_\mu (C_V^s -C_A^s \gamma_5) s),
\eea where $C_{V,A}^s$ are the vector/axial vector couplings of $s
\bar s$ quarks to the $Z$ boson.  Now to evaluate the transition
amplitude, we use the following matrix element \bea \langle \phi(p',
\epsilon) | \bar s \gamma^\mu s |0 \rangle = f_\phi m_\phi
\epsilon^\mu \eea where $f_{\phi}$ is the decay constant of $\phi$
meson and $\epsilon $ being its polarization vector. With this we
obtain the amplitude for this process as \bea {\cal M}(\tau \to \mu
\phi)=\frac{G_F}{\sqrt 2} U_{\mu \tau} C_V^s f_\phi m_\phi (\bar \mu
\gamma^\mu (1-\gamma_5) \tau) \epsilon_\mu . \eea Neglecting the
muon mass, the corresponding decay rate is found to be \bea \Gamma =
\frac{G_F^2}{16 \pi} |U_{\tau \mu}|^2 (C_V^s)^2 f_\phi^2 m_\phi^2
m_\tau \left (1 - \frac{m_\phi^2}{m_{\tau^2} }\right )^2 \left ( 2+
\frac{m_\tau^2}{m_{\phi}^2} \right ) \eea Now using  $\phi$ meson
decay constant $f_\phi=0.231$ GeV, and varying $U_{\tau \mu}$
between $0\leq |U_{\tau \mu}| \leq 1.295 \times 10^{-3}$, we present
in Figure-6, the correlation plot between $\tau \to \mu \mu \mu$ and
$\tau \to \mu \phi$. Since the upper limit of ${\rm Br}(\tau \to \mu
\phi) \sim {\cal O} (10^{-8})$, there is possibility that this mode
could be observable in the near future.

 \begin{figure}[htb]
  \centerline{\epsfysize 2.25 truein \epsfbox{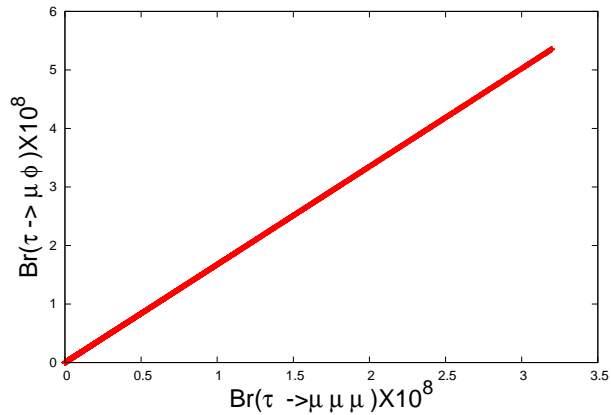}}
\caption{Correlation plot between the branching ratios of $\tau^-
\to \mu^- \mu^+ \mu^+ $ and $\tau \to \mu \phi$ processes.}
  \end{figure}

\section{$\mu - e$ Conversion in Nuclei and $\mu \to e \gamma$ process}

Now we will consider another example of lepton flavour violating
process i.e., conversion of muon into electron in nuclei, where very
stringent experimental upper bounds exist.  In particular, the
experimental upper bound on $\mu -e$ conversion in $_{22}^{48}Ti$ is
\cite{mue1} \be R(\mu Ti \to e Ti) < 4.3 \times 10^{-12},\ee and the
dedicated J-PARC experiment PRISM/PRIME \cite{mue2} should reach a
sensitivity of ${\cal O}(10^{-18})$. A very detailed calculation of
$\mu-e$ conversion rate in various nuclei has been performed in
\cite{mue3}. Following Ref. \cite{mue4},  one can obtain the
conversion rate in nuclei (normalized to the total
nuclear muon rate $\Gamma_{capture}$) in the FCNC mediated $Z$ boson model as
 \be R (\mu Ti \to e Ti)= \frac{G_F^2 \alpha^3 m_\mu^5}{4\pi^2} \frac{Z_{eff}^4}{Z}
\frac{|U_{\mu e}|^2 }{\Gamma_{capture}}|F(q^2)|^2 Q_W^2 \ee where
\be Q_W= (2 Z+N)C_V^u+(Z+2N)C_V^d, \ee is the coherent nuclear
charge associated with the vector current of the nucleus, which
gives an enhanced contribution to the coherent nuclear transition.
$C_V^u$ and $C_V^d$ are the vector couplings of up and down quarks
to the $Z$ boson given as \be C_V^u= \frac{1}{2} -\frac{4}{3} \sin^2
\theta_W,~~~~~~~C_V^d= -\frac{1}{2} +\frac{2}{3} \sin^2 \theta_W.
\ee $F(q^2)$ is the nuclear form factor and for $_{22}^{48}Ti$ its
value is found to be $F(q^2=-m_\mu^2)\simeq 0.54 $, and $Z_{eff} \simeq
17.6 $ \cite{mue4}. For $\Gamma_{capture}$, we use its experimental
value $\Gamma_{capture}=(2.590 \pm 0.012)\times 10^6 sec^{-1}$ \cite{mue5}.
The variation of $R(\mu Ti \to e Ti)$ with $|U_{\mu e}|$
is shown in Figure-7 (red curve). The horizontal line represents the experimental
upper limit. From the figure it can be seen that much stronger constraint
on $|U_{\mu e}|$, i.e., $|U_{\mu e}| < 1.05 \times 10^{-6}$,
can be obtained from the $\mu -e$ conversion rate.
This bound on $|U_{\mu e}|$ is about three times stronger
than the constraint obtained from $\mu \to eee$
process.
 \begin{figure}[htb]
  \centerline{\epsfysize 2.0 truein \epsfbox{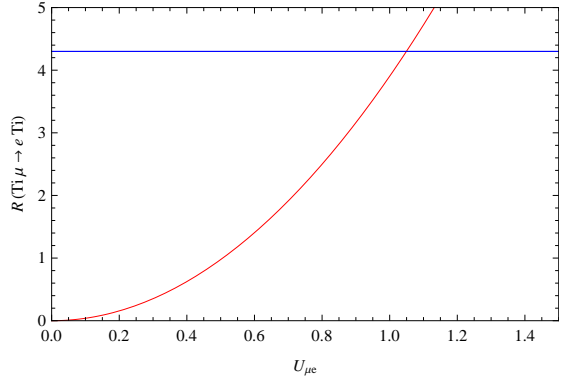}}
\caption{Variation of $\mu-e$ conversion rate $R(\mu Ti \to e Ti)$ (in units
  of $10^{-12}$) with $|U_{\mu e}|$ (red curve). The horizontal blue
   line represents the experimental upper limit.}
  \end{figure}


Another well-known example of lepton flavour violating process is
$\mu \to e \gamma $. However, this process occurs at one loop level
in the FCNC mediated $Z$ boson model as shown in Figure-8, where the
internal fermion line $l_i=(e,\mu,\tau)$. Here we will consider the
internal fermion lines to be either $\mu$ or $e$ so that we will have
only one FCNC  $Z\mu e$ vertex. However, we will neglect the contribution
coming from internal electron line as it is proportional to $(m_e/m_\mu)$.
Thus, we obtain the decay rate for $\mu \to e \gamma $ as
\be \Gamma(\mu \to e \gamma) =
\frac{\alpha G_F^2 m_\mu^5 }{32 \pi^4} |U_{\mu e}|^2 (C_V^\mu - C_A^\mu)^2\;.
\ee
Now using the bound on $|U_{\mu e}| < 10^{-6}$, we obtain the branching ratio
\be
\rm{Br}(\mu \to e \gamma) < 3 \times 10^{-15},
\ee
which is well below the present experimental upper limit
\cite{pdg}
\be
\rm{Br}(\mu \to e \gamma)_{expt} < 1.2 \times 10^{-11}.
\ee

 \hspace*{1.0 truein}
\begin{figure}[b]
\begin{center}
\begin{picture}(300,100)(0,0)
\SetColor{Black} \PhotonArc(150,50)(50,0,180){5}{7} \SetColor{Black}
\ArrowLine(50,50)(100,50){}{} \Vertex(100,50){2} \SetColor{Black}
\ArrowLine(200,50)(250,50){}{}\Vertex(200,50){2}
\ArrowLine(100,50)(200,50){}{} \Photon(155,50)(180,15){5}{5}
\Text(135,40)[]{$l_i$} \Text(75,40)[]{$\mu$} \Text(150,115)[]{$Z$}
\Text(225,40)[]{$e$} \Text(190,06)[]{$\gamma$}
\end{picture}
\end{center}
\caption{Feynman diagram
 for $\mu \to e \gamma $ in the model with FCNC mediated
 $Z$ boson.}
\end{figure}
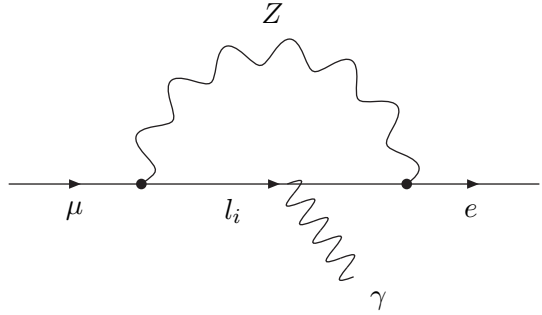
\begin{table}[t]
\begin{center}

\begin{tabular}{|c |c|c|}
\hline \hline
Decay process & Predicted Br & Experimental \\
 & (maximal value) & upper limits \cite{pdg} \\
\hline $\mu^- \to e^- e^+ e^-$ & $1.0 \times 10^{-12}$ & $1.0 \times
10^{-12}$\\

$\tau^- \to e^- \mu^+ \mu^- $ & $3.7 \times 10^{-8} $& $3.7 \times
10^{-8}$\\
$\tau^- \to \mu^- \mu^+ \mu^- $ & $3.2 \times 10^{-8} $ & $3.2
\times
10^{-8}$\\
\hline

 $\mu^- \to e^- \nu_e \bar \nu_\mu $ & $ 4.36 \times
10^{-23}$ & $
1.2 \times 10^{-2}$\\

$\tau^- \to e^- e^+ e^- $ & $ 3.1 \times 10^{-8}$ & $ 3.7 \times
10^{-7}$\\
$\tau^- \to \mu^- \phi $ & $ 5.4 \times 10^{-8}$ & $ 1.3 \times
10^{-7}$\\

$B_d \to e^\pm \mu^\mp $ & $ 1.7 \times 10^{-19}$ & $6.4 \times
10^{-8}$\\
$B_s \to e^\pm \mu^\mp $ & $ 4.7 \times 10^{-18}$ & $2.0 \times
10^{-7}$\\
$B_d \to e^\pm \tau^\mp $ & $ 2.1 \times 10^{-10}$ & $2.8 \times
10^{-5}$\\
$B_s \to e^\pm \tau^\mp  $ & $ 1.9 \times 10^{-10}$ & $-$\\
 $B_d \to \mu^\pm \tau^\mp $ & $ 2.1 \times 10^{-10}$ & $2.2 \times
10^{-5}$\\
 $B_s \to \mu^\pm \tau^\mp $ & $ 1.9 \times 10^{-10}$ & $-$\\
 \hline
 \hline
\end{tabular}
\caption{ Maximal values of the branching ratios for LFV  decays in
the FCNC mediated  $Z$ boson model, after imposing the constraints
on ${\rm Br }(\mu^- \to e^- e^+ e^-$,  $\tau^- \to e^- \mu^+ \mu^-$
and $\tau^- \to \mu^+ \mu^- \mu^-$).}
\end{center}
\end{table}

\section{Conclusion}

We have studied LFV decays of $\mu$, $\tau$ and $B_{d,s}$ mesons in
the model with additional vector-like leptons. In such a model due
to the mixing between the exotic singlet leptons with the SM
leptons, flavor changing neutral current transitions can occur at
the tree level  mediated by $Z$ boson. Due to such couplings, the
LFV decay modes considered here $l_i \to l_j l_k \bar l_k$, $B_{d,s}
\to l_i \bar l_j$, $\tau \to \mu \phi$ and the $\mu-e$ conversion
in nuclei can arise at the tree
level in this model. Assuming that the SM contributions to such
decay modes have negligible effect we obtain the branching ratios
for these LFV modes. The constraint on the new physics parameters
are obtained using the present experimental limits on $\mu^- \to e^-
e^+ e^-$, $\tau^- \to e^- \mu^+ \mu^-$ and $\tau^- \to \mu^- \mu^+
\mu^-$. These bounds impose strong constraints on the branching
ratios of the other LFV decays. In Table-1, we present the upper
limits of the branching ratios of various LFV decay modes using
these constraints. We find that the LFV decays involving a $\tau$
meson i.e., ${\rm Br }(B_{d,s} \to \tau^\pm \mu^\mp, \tau^\pm e^\mp)
\sim {\cal O} (10^{-10})$, which could be observed in the upcoming
high sensitivity experiments. However, analogous LFV decays such as
$B_{d,s} \to \mu e$, have branching ratios of the order of ${\cal
O}(10^{-18})$, and  are too small to be observed in the near future.
We also find that the branching ratio of the semileptonic decay of
$\tau$ meson $\tau \to \mu \phi$  could be as large as ${\cal
O}(10^{-8})$ in this model, the observation of which would
unambiguously point to the presence of new physics.
However, the branching ratio for $\mu \to e \gamma$ is found to be 
quite small in this model as it occurs at one-loop level.

 {\bf Acknowledgments}

The author would like to thank Department of Science and Technology,
Government of India, for financial support through Grant No.
SR/S2/RFPS-03/2006.


\end{document}